\documentclass[conference]{IEEEtran}
\IEEEoverridecommandlockouts
\usepackage{amsmath,amssymb,amsfonts}
\usepackage[dvipdfmx]{graphicx}
\usepackage{textcomp}
\usepackage{xcolor}
\usepackage[normalem]{ulem}
\usepackage[style=ieee]{biblatex}
\usepackage{algorithm}
\usepackage{algpseudocode}
\AtBeginBibliography{\small}
\def\BibTeX{{\rm B\kern-.05em{\sc i\kern-.025em b}\kern-.08em
    T\kern-.1667em\lower.7ex\hbox{E}\kern-.125emX}}
\begin{document}

\title{IoT Security and Safety Testing Toolkits for Water Distribution Systems
\thanks{Funding provided under a grant from the National Science Foundation.}
}
\author{
 \IEEEauthorblockN{Sean O'Toole, Cameron Sewell, Hoda Mehrpouyan}
 \IEEEauthorblockA{\textit{Department of Computer Science} \\
 \textit{Boise State University}\\
 Boise, United States \\
 seanotoole@u.boisestate.edu, cameronsewell@u.boisestate.edu, hodamehrpouyan@boisestate.edu}
 }
\maketitle

\begin{abstract}
Due to the critical importance of Industrial Control Systems (ICS) to the operations of cities and countries, research into the security of critical infrastructure has become increasingly relevant and necessary. As a component of both the research and application sides of smart city development, accurate and precise modeling, simulation, and verification are key parts of a robust design and development tools that provide critical assistance in the prevention, detection, and recovery from abnormal behavior in the sensors, controllers, and actuators which make up a modern ICS system. However, while these tools have potential, there is currently a need for helper-tools to assist with their setup and configuration, if they are to be utilized widely.  Existing state-of-the-art tools are often technically complex and difficult to customize for any given IoT/ICS processes. This is a serious barrier to entry for most technicians, engineers, researchers, and smart city planners, while slowing down the critical aspects of safety and security verification. To remedy this issue, we take a case study of existing simulation toolkits within the field of water management and expand on existing tools and algorithms with simplistic automated retrieval functionality using a much more in-depth and usable customization interface to accelerate simulation scenario design and implementation, allowing for customization of the cyber-physical network infrastructure and cyber attack scenarios. We additionally provide a novel in-tool-assessment of network's resilience according to graph theory path diversity. Further, we lay out a roadmap for future development and application of the proposed tool, including expansions on resiliency and potential vulnerability model checking, and discuss applications of our work to other fields relevant to the design and operation of smart cities. 
\end{abstract}


\section{Introduction}
As a result of general increases in malicious cyber activity in recent years, and in the wake of STUXNET, the cyber security of industrial control systems is increasingly becoming a massive security concern for cities, companies, and nations \cite{dragos_year_2020, slowik_evolution}.  The most prominent types of attacks have usually targeted the Information Technology (IT) components of a system. These attacks are usually conducted for financial gain and usually do not cause long term damage to the systems being affected, especially if there is a significant separation between data management and the actual operational technology (OT). For example, the recent Colonial Pipeline attacks, which were IT-based and targeted billing information. Colonial themselves shut down their own pipeline until the ransomware issue was resolved and there was no structural damage done to the pipeline \cite{jun_7_colonial_2021}. If, however, they were targeted with an OT-focused attack that took control of operational devices, sensors, or actuators within the ICS, the attackers could have created serious physical damage in the over 5,500-mile-long pipeline that could have caused millions of dollars damage and left up to 45 percent of the East Coast of the United States without gas for an undetermined amount of time. 

An example of the full potential damage of an OT attack could be seen in the 2007 INL Aurora Generator test \cite{muckrock_aurora}, which created abnormal vibrations in a diesel generator within 13 iterations of the attack loop and eventually caused the engine to dramatically fail. Manipulation of the control logic which controlled the generator led to significant physical damage, a methodology that would be similarly applied in the cyber worm STUXNET several years later in an attack against the country of Iran's nuclear refinement capabilities to significant effect. 

In the years since, an increasing number of attacks have sought to utilize tools like this against city and regional infrastructure. In 2015, several regions of Ukraine’s energy grid were affected by an attack, leaving 225,000 people without power. And again in 2016 a power grid hack affecting a key substation shut down one fifth of power traffic in the city of Kyiv. Following the 2016 attack the President of Ukraine, Petro Poroshenko, stated that state institutions had been targeted about 6,500 times in the final months of that year~\cite{ukraine_2017}. 2017's TRITON attacks against Saudi Arabian natural gas processing plants represented a near-miss, where the potential for serious physical damage and threat to human life was only avoided by a single fail-safe missed by attackers~\cite{futureiot_triton}.  In regards to water distribution systems, the Oldsmar Water Plant Attack in Feb. 2021 is the most recent example \cite{fedtech_cybersecurity_nodate}, following on the heels of the April 2020 attacks against multiple regional water systems in Israel \cite{cimpanu_two_israel}, both of which sought to dump toxic levels of chlorine into the water, presenting a serious threat to the health and well being of citizens. However, OT threats to water go back further, all the way to the 2000 Maroochy water plant incident, in which hundred of thousands of gallons of untreated sewage were dumped back into local water supplies by a single attacker with access to OT systems \cite{abrams_malicious_maroochy}. For a more thorough review of cyber attacks against ICS in general, we refer the reader to an excellent review of incidents up to 2018 in Hemsley et al. \cite{hemsley_history_2018}. To better protect our critical infrastructure we need to understand exactly how these OT attacks affect our infrastructure so that we can create safe guards to protect against them. We additionally need to develop ways to better prepare and train the professionals in the field for incidents such as these. For this, accessible and usable tools are badly needed.

The rest of the paper is organized as follows: Section 2 provides context for the application of simulation in various fields for the purpose of improving safety and security in systems and procedures, as well as more specifically those targeting the needs of smart cities. Section 3 describes state-of-the-art simulation toolboxes for Water Distribution Systems (WDS) which could be used for research and practical testing of hydraulic systems models. Section 4 introduces \textit{inp2cpa} ~\cite{inp2cpa_source} and the changes we have made to the tool in order to make file conversion and creation more accessible and streamlined, as well as our contribution of network graph theory resiliency model checking, and briefly discusses future plans for the work, which is expanded upon in Section 5. Finally, Section 5 summarizes the work, and more thoroughly outlines future plans and goals for \textit{inp2cpa} moving forwards. 

\section{Simulation for Improving Safety and Security}
Simulation here refers to the replication of real-world processes with cyber and cyber-physical tools and programs by means of mathematical equations and programmatic structures which capture, as closely as possible, the key characteristics of the system being simulated. We will address some of the many ways in which simulation has been used to improve safety and security outcomes, and then focus in on open-source toolkits for smart city IoT simulation applications in order to identify those which would be useful for cities to test and improve on their systems without needing to rely on or expend heavily into proprietary software.

Many fields have incorporated simulation into their education and training to yield well-educated individuals at lower cost and less danger to the individual in-the-field training. The precursor to modern simulation training was first tested in 1978 by the National Aeronautics and Space Administration, who found in their studies of disasters that most aviation accidents involved a lack of leadership coordination, or decision making~\cite{carron2011high}. Simulations also spread to the maritime world focusing on on the same issues. Most high fidelity maritime simulations reproduce the pilot or captains room and the engine room to inundate the crew with the auditory and visual factors they will need to expect while on the job~\cite{a0958ff56e4f442ba1a114fd3db062f0, helmreich2001improving,HETHERINGTON2006401}. The formalization of simulation-based educational practices has advanced in the medical industry as well due to the development of international and national multidisciplinary societies, usually aimed at a broad membership of healthcare educators, clinicians, researchers and engineers, and supported by industry \cite{https://doi.org/10.1111/nicc.12030}. These simulation training sessions better prepare individuals for the needs of different scenarios than standard training, reducing accidents and waste. In all cases, simulation could be applied to improve safety outcomes for both those doing the work and those relying on their efforts.

Within those subset of simulators targeting the needs of smart cities and considering the role of IoT devices within those cities, and which are seeing ongoing development and maintenance, several exist which allow for relatively thorough simulation of various network, power, and IoT simulation. Simulation of Urban Mobility (SUMO) is one of the oldest and best examples of the kind of toolset that can be of great benefit to a city, offering a powerful tool for analyzing road networks and traffic patterns and experimenting on them, before returning to users simulated traffic data, physical effect data, and improved road networks if  desired by the user~\cite{sumo_bisw_international}.

Bonceur's CupCarbon~\cite{cupcarbon} is perhaps the best example of a modern smart city simulation tool with the capability of modeling real-world scenarios and their impacts on and interactions with a specific subset of city infrastructure. Bounceur proposed and later presented CupCarbon in 2017 as a flexible simulation tool focused on radio propagation channels and IoT interactions within a user-designated network and set against real-world maps. CupCarbon allows for the simulation of wireless network behavior and communication with a large number of different components and customizable parameters, and most notably allows for designing scenarios in which real-world events such as fire incidents and gas leaks can be simulated, with the event having an effect on the ability of the network to communicate clearly. Custom python scripts can be written for different nodes, allowing for potentially limitless variety and customization of system behavior. However, with few examples and no capability of importing existing files on which to base designs, the tool is currently inhibited by its lack of user accessibility and the lack of training available for it.

It is our goal with this work to help streamline the process of simulation in industrial control system security and infrastructure resilience~\cite{}, as well as bring better physical systems simulation and OT more into focus when discussing smart city infrastructure and security.

\begin{figure}[htbp]
    \centering
    \includegraphics[width=7cm]{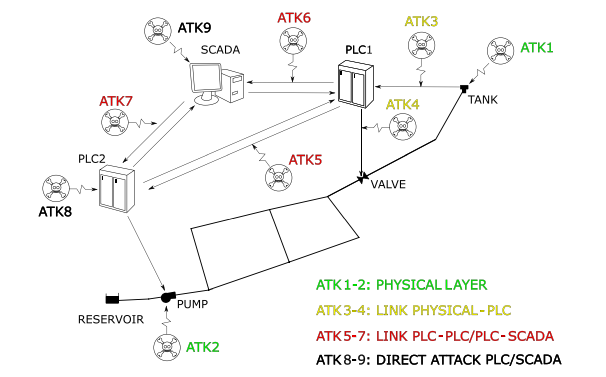}
    \caption{epanetCPA Attack Modeling Approach - "Characterizing Cyber-Physical Attacks on Water Distribution Systems," Toarmina et al. \cite{taormina2017characterizing}}
    \label{fig:epanet}
    \vspace{-10pt}
\end{figure}

\begin{figure*}[htbp]
    \centering
    \includegraphics[width=12cm]{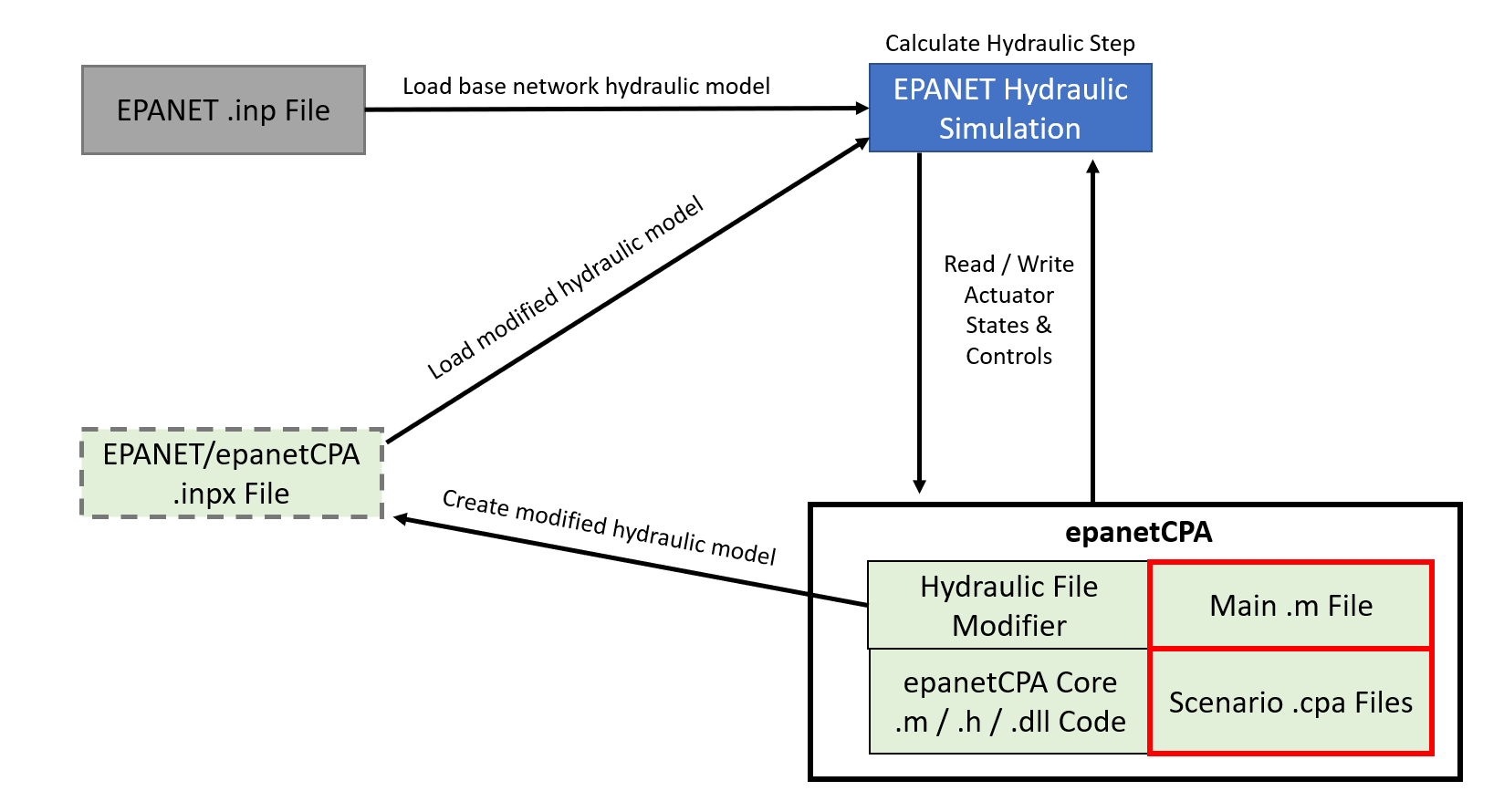}
    \caption{File requirements for epanetCPA configuration. Red indicates user creation required.}
    \label{fig:epa_config}
\end{figure*}
\begin{figure*}[htbp]
    \centering
    \includegraphics[width=10cm]{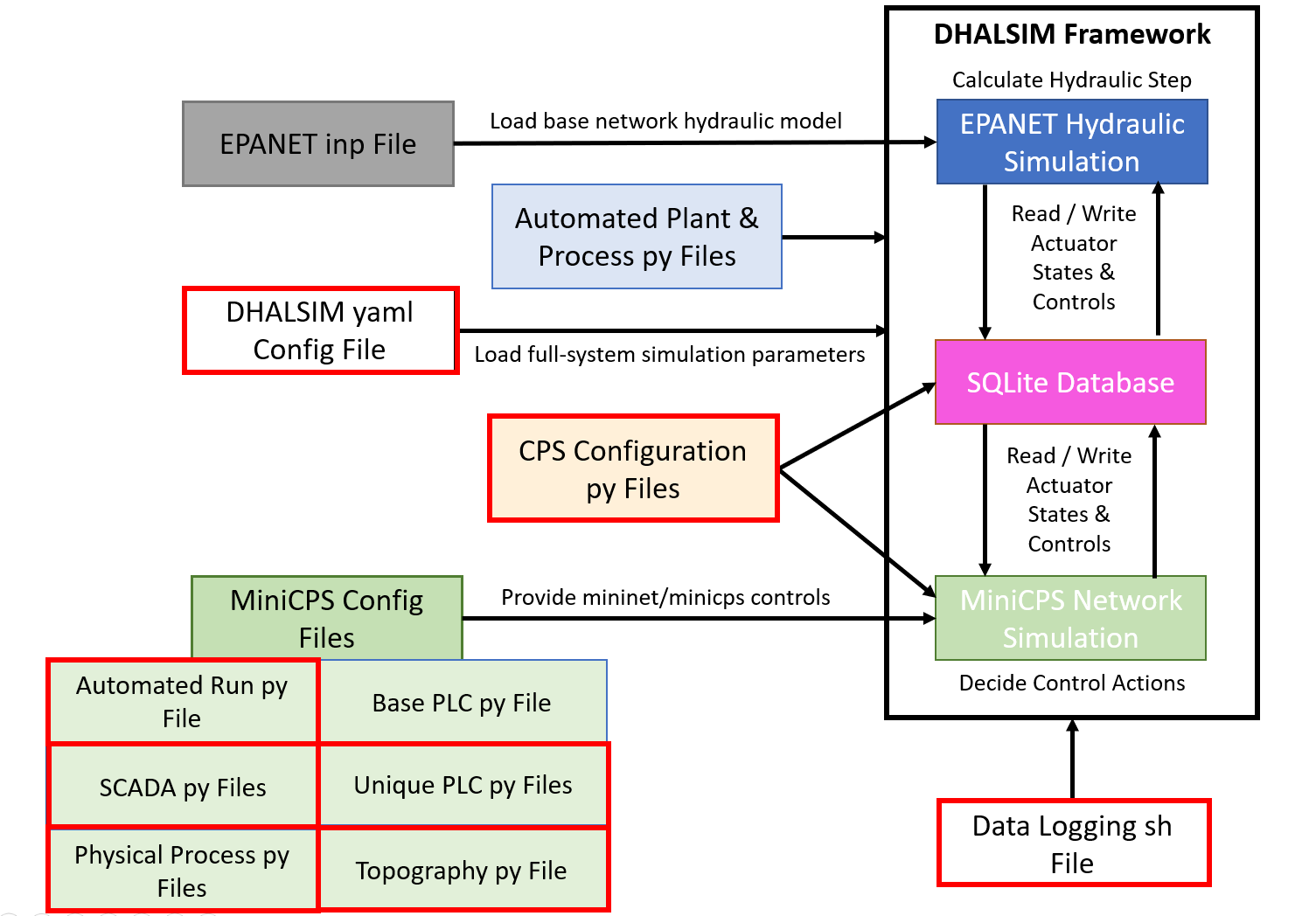}
    \caption{File requirements for DHALSIM configuration. Red indicates user creation required.}
    \label{fig:dhal_config}
    \vspace{-10pt}
\end{figure*}

\section{State-of-the-Art Safety and Security Testing Toolboxes in WDS}
A variety of open-source tools have been created to address the threat of OT attacks, specifically in water distribution. Among these, the two we find to be most significant in terms of contribution and potential for future development in recent years are Taormina et al.'s epanetCPA \cite{taormina2017characterizing} and Murillo et al's. DHALSIM  \cite{murillo_co-simulating_2020}. Taormina et al. and his team created epanetCPA, a MATLAB based program for modeling the response of water systems to cyber-physical attacks that runs on top of EPANET. EPANET is used world wide by engineers and researches to design new water infrastructure, update existing water systems, and develop more efficient solutions to solve water quality problems. It is a Windows based software application for simulating and representing water distribution systems. It uses a network configuration or .inp file which represents a physical water system and creates a digital representation of the water network, representing all pipes, junctions, tanks, reservoirs, valves, and pumps. epanetCPA works in tandem with EPANET modifying the input file with information from a user provided cyber-physical attack file, .cpa. The changes in file information create a new data set seperate from the ground truth and diplayed by the EPANET model analysis. Murillo et al. in "Co-Simulating Physical Processes and Network Data for High-Fidelity Cyber-Security Experiments" introduced Digital Hydraulic Simulator (DHALSIM) ~\cite{murillo2020co} to show the physical processes, control logic, and network communication of a cyber-physical system while it is under attack. DHALSIM provides users a full network capture of the PLCs, SCADA systems, and any other network device present on the system.

However, although both toolkits offer considerable utility, creating custom scenarios and models for each requires considerable work on the part of the user, and the prerequisite knowledge of the programming languages and/or file formats in use within each toolkit. As can be seen in Fig. \ref{fig:epa_config} and Fig. \ref{fig:dhal_config}, multiple files of different formats and languages are required for each of epanetCPA and DHALSIM, although the added complexity of DHALSIM in attempting to co-simulate network behavior results in requiring significantly many more custom files and configurations. This is a non-trivial issue for research and application of these toolkits going forward, if they are to be utilized for safety and security testing and training. While some understanding of programming is not an unreasonable expectation for users, the current toolkits require an in-depth understanding of the toolkits themselves, along with the formats and langauges in use, in order to customize to any significant degree. Cutting down on this requirement would make them vastly more approachable and usable. We believe that \textit{inp2cpa} and expansion of its current functionality, and the creation of tools like it for other simulation toolkits in WDS and general ICS simulation, is a critical part of lowering that barrier to entry.

\section{\textit{inp2cpa}: A Tool for Rapid Integration}
\subsection{\textit{inp2cpa} Basic Functionalities}
As mentioned, the epanetCPA toolkit requires an engineer to have knowledge on how to extract relevant data from .inp files and use that information to create .cpa files. Learning these file types and their syntax can be difficult and time consuming to correctly implement for those without experience with relevant field-specific programs, as well as the toolkits themselves. To make these toolkits more accessible we have taken and extended the functionalities of the tool \textit{inp2cpa}, created by Nikolopuolos~\cite{inp2cpa_source, nikolopoulos_cyber-physical_2020} and the EU STOP-IT team, to drastically reduce the time it takes to run bulk test simulations as well as create custom cyber-physical attacks to simulate on a water network. This tool is a python-based program that utilizes the WNTR toolkit \cite{klise2018overview}. WNTR is a python package created by the EPA and Sandia Labs to evaluate  drinking water systems for water quality and resilience against damgage cause by demand, natural disasters, and cyber attacks. WNTR is capable of generating water network models, modifying network structure, evaluating disruptive events, simulating pressure driven and demand driven hydraulics. It is used by engineers in the infrastructure industry as well as other cybersecurity researchers around the world.
\begin{figure}[htbp]
    \centering
    \includegraphics[width=7cm]{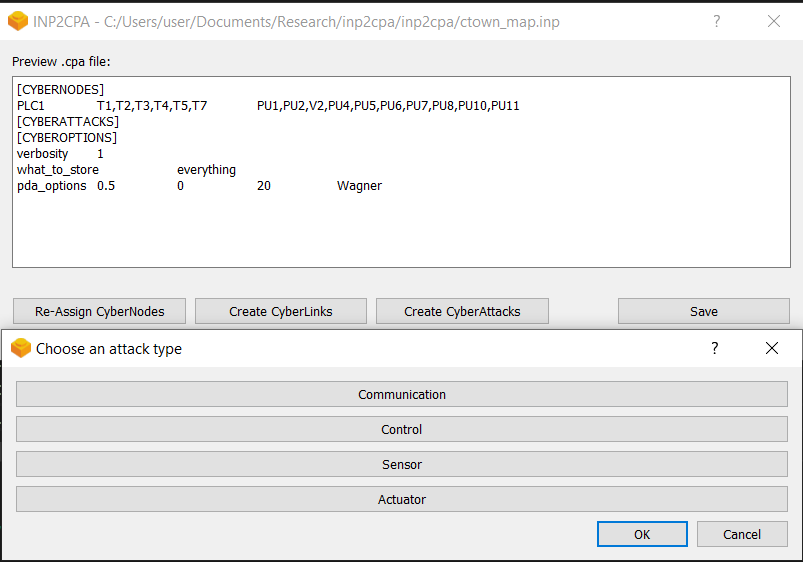}
    \caption{\textit{inp2cpa} Base Menu and Attack Selection Window}
    \label{fig:inpmain}
    \vspace{-10pt}
\end{figure}
\begin{figure}[htbp]
    \centering
    \includegraphics[width=7cm]{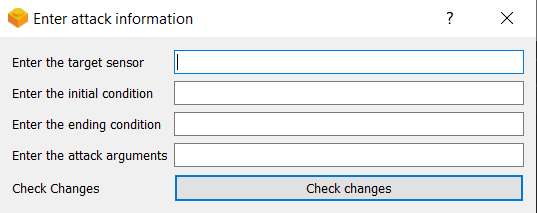}
    \caption{\textit{inp2cpa} Sensor Attack Specification Window}
    \label{fig:inpsens}
    \vspace{-10pt}
\end{figure}
\textit{inp2cpa} uses WNTRs file parsing capabilities in its network.controls script to build a python dictionary using  the control section of .inp files. With this dictionary \textit{inp2cpa} autmoatically generates a basic .cpa file. We have expanded \textit{inp2cpa} to allow users to modify the automatically generated cyber-attack file by adding Cyberlinks and Cyberattacks. A Cyberlink is a way that a user can show how their hardware is connected. For example, if a series of actuators where connected to one plc, a user could add cyberlinks to the .cpa show the link between each actuator and its plc and then the plc to its SCADA system. Adding these links will allow the user to simulate attacks that  take advantage of or disrupt those connections. The user can now add four types of cyber-attacks:
\begin{itemize}
    \item \textit{Communication} - intercepting and replacing communications between hardware
    \item \textit{Control} - Changing the control logic of how physical components operate
    \item  \textit{Sensor} - manipulates the values sent out by sensors
    \item \textit{Actuator} - take control of a physical component (e.g. valves, pumps) and manipulate it
\end{itemize} 
We have simplified the process of creating these attacks by getting key information like the names of the sensor to be attacked and what values the want to interject into that sensor and then using string manipulation and concatenation we add it to a .cpa file with the correct syntax (as specified in Munteanu et al. \cite{TAORMINA201946})  and create an attack file ready for use. \textit{inp2cpa} is meant to be used specifically with epanetCPA, and our contribution has expanded on the original automated processing of controls by completing several of the originally-proposed features, including customization and control menus, as well as numerous fixes to the text-parsing and error-checking features of the original code. We intend, however to make it compatible with DHALSIM as well by adding YAML, “Yet Another MarkUp language", used for DHALSIM configuration files, as an output, along with potentially other languages for file conversion and creation.\\

\begin{figure*}[htbp]
    \centering
    \includegraphics[width=8cm]{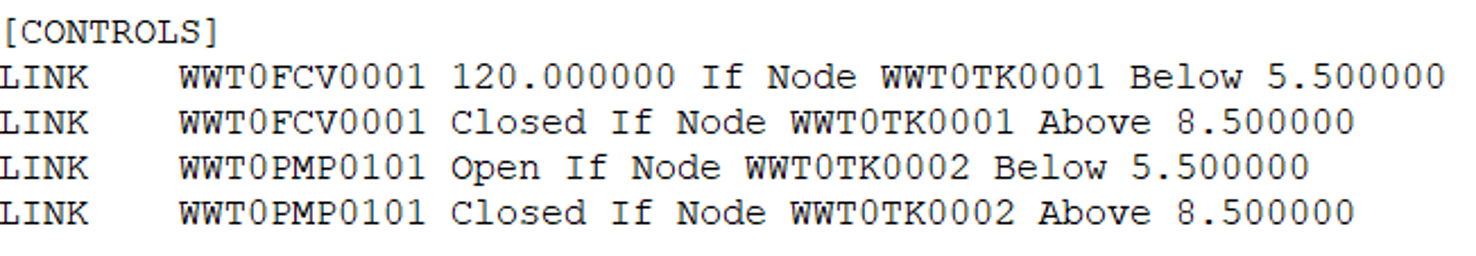}
    \caption{Control section of an .inp file}
    \label{fig:inp}
\end{figure*}
\vspace{-20pt}

\begin{figure*}[htbp]
    \centering
    \includegraphics[width=8.5cm]{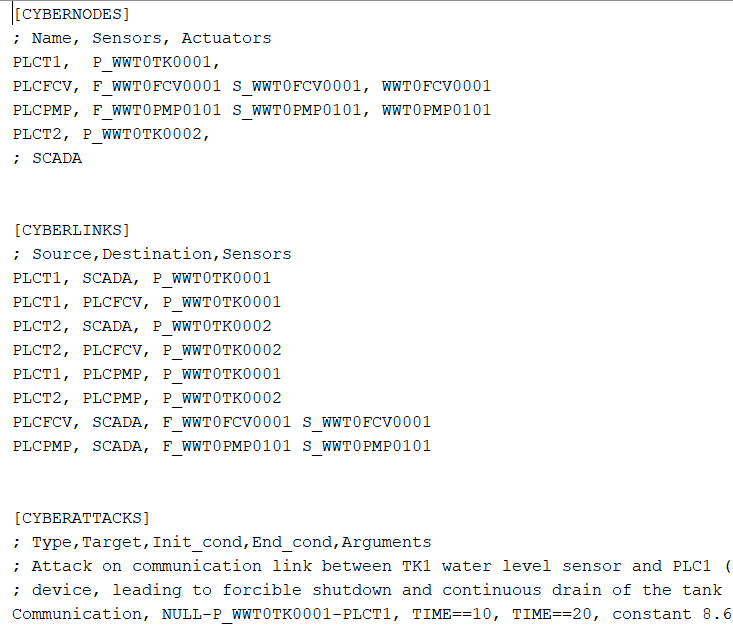}
    \caption{Post-processing of Fig. 4 within \textit{inp2cpa}, resulting .cpa file, including user-specified attack against communication link.}
    \label{fig:cpa}
    \vspace{-10pt}
\end{figure*}

\vspace{10pt}
\subsection{\textit{inp2cpa}: Resiliency Checking On-the-Fly}
\begin{figure}[htbp]
    \centering
    \includegraphics[width=7cm]{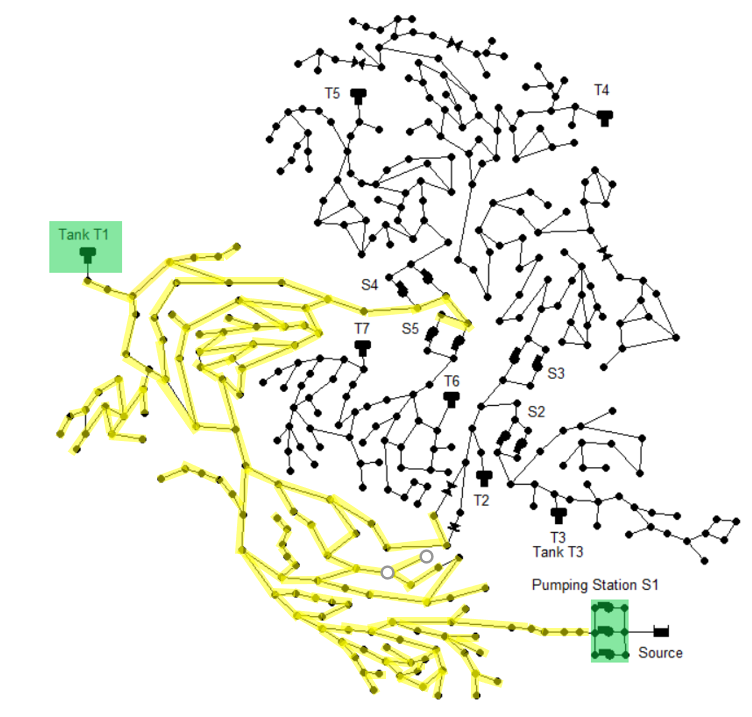}
    \caption{Subsection of CTown Relevant to DoS Attack}
    \label{fig:DoS_map}
    \vspace{-10pt}
\end{figure}

In order to allow in-application review of the user's networks against common network resiliency metrics, \textit{inp2cpa} creates a \textit{G(V,E)} graph theory model of the network as defined by the user, using the general form of: \begin{gather*}
    G\{Nd, Ln\} \\
    Nd \{ S, A, Con \} \\
    Ln\{So, De, S\}
\end{gather*}
, wherein a network graph is modeled by a graph \textit{G} of nodes \textit{Nd} and links \textit{Ln}, each of which is represented by a set of tuples representing \textit{Nd\{\textbf{S}ensors, \textbf{A}ctuators, \textbf{Con}trols}\} and \textit{Ln\{\textbf{So}urce, \textbf{De}stination, \textbf{S}ensors}\}. Utilizing this approach, a number of network resiliency metrics can be tested against, to give immediate feedback to users on their network's likely resilience against disruption and/or malicious activity. 
As a case study for this, we utilized automatic input of the common hydraulic simulation baseline of the \textit{ctown} model alongside user-defined custom nodes and links, extracting a subset of the hydraulic network representing the control nodes PLC1 and PLC2, responsible for respectively: monitoring pump flow and junction pressure on the first three pumps in the network; monitoring water reserve tank level for the corresponding pumped reserve tank. This subset of the overall network represents just under half of the overall network capacity and complexity, and can be seen in in Fig.\ref{fig:DoS_map} However, while it represents a high level of physical complexity, the logical complexity is often significantly lower. As shown in Taormina et al. \cite{taormina2017characterizing} in their cybernode and link layouts for the ctown attack scenarios, a common way this might be handled is simply by assigning one PLC to monitor tank levels and separately, one to control the pumps feeding the network and monitor pressure levels on either side of the pumps for potential issues. However, as demonstrated by their own attack simulations against this set of control components, such a simple design has effectively no resiliency against any attack which interrupts or replaces valid communication between components, nor did they attempt to quantify or otherwise assess the resilience of cyber components of the network. 

\begin{figure}[htbp]
    \centering
    \includegraphics[width=7cm]{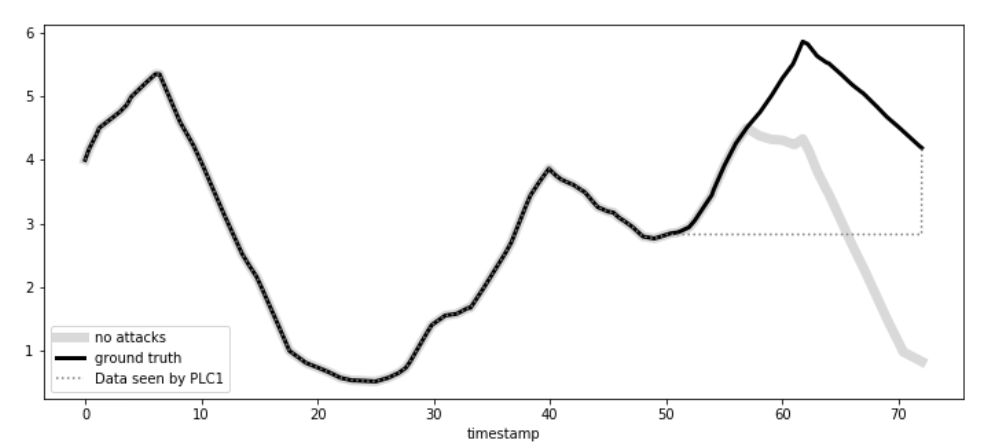}
    \caption{Denial-of-Service Attack against PLC-monitored tank system}
    \label{fig:DoS_unmod}
    \vspace{-10pt}
\end{figure}

The results of this attack as simulated in Fig. \ref{fig:DoS_unmod} show that even interrupting one communication link between the sensor-monitoring PLC and the pump-controlling PLC will result in the system being entirely unable to identify and respond to conditions in the tank. While pumps should have been disconnected at \~58 hours, the pumps continue to run and create overflow. Either duplicate tracking on the relevant sensor by the pump PLC or additional nodes (for example, a SCADA unit) tracking and sending data to nodes waiting on information would have resolved this, but the total lack of either results in likely damage to the system. While total duplication and redundancy across all nodes and controls in a network is impractical financially, and introduces both considerable extra complexity and potential additional surface area for attackers to target, we believe that even a small amount of additional connectivity and built-in resiliency in design would contribute significant value here, if only by ensuring that multiple attacks or interruptions would need to occur simultaneously to guarantee significant impact on the system.

In addressing this gap, we utilized the aforementioned graph theory model ~\cite{mehrpouyan2012model,mehrpouyan2015resiliency,mehrpouyan2013resilient} alongside the findings of Alenazi et al. \cite{alenazi_evaluation_2015}, in which Total Graph Diversity (TGD) \ref{alg:tgd} was identified across multiple structured and random network designs to be the most effective predictor of network resiliency against outages and directed attacks, to evaluate different configurations of the cyber-physical network. This metric, which is first described in Rohrer et al. \cite{rohrer_path_2014} is measured as the average of Effective Path Diversity (EPD) \ref{eq:epd}, representing lowest path diversity \ref{eq:pdiv}  over a set of paths through the network, in this case over the directed edge links from node to node. Path diversity is determined for any given path among a set of possible paths between two nodes by a simple equation evaluating path length against shortest possible path length. EPD additionally factors in a $\lambda$ value which is decided upon for any given network based on experimentation and desired outcomes. As a result of the exponential nature of the EPD formula, greater $\lambda$ values correlate with a reduced impact of $K_{sd}$ (minimum path diversity between two points in a graph \ref{eq:ksd}) values on the outcome, indicating reduced gains on resiliency for additional path diversity. Pseudocode for a simple implementation of each can be found in \ref{alg:tgd} and \ref{alg:epd}. 

\begin{equation}
    D(P_a,P_b)=1-\dfrac{|P_a \oplus P_b|}{|P_b|}\label{eq:pdiv}
\end{equation}   
\begin{equation}
     EPD(P_a,P_b)=1-\exp^{-\lambda*K_{sd}} \label{eq:epd}
\end{equation}
\begin{equation}
     K_{sd}=\sum_{i=1}^{k}{D_{min}(P_i)} \label{eq:ksd}
\end{equation}

\begin{algorithm}
\caption{TGD($Graph$):}\label{alg:tgd}
\begin{algorithmic}
    \State $ sum \gets 0 $
    \State $ cnt \gets 0 $
        \For {$node_s$ \textbf{in} $Graph$}
            \For {$node_d$ \textbf{in} $(Graph - node_s)$} 
                \State $ cnt \gets cnt + 1$ 
                \State $ sum\gets sum +  EPD(Graph,node\_s,node\_d) $
            \EndFor
        \EndFor
    \State \textbf{return} $sum/cnt$
\end{algorithmic}
\end{algorithm}
\begin{algorithm}
\caption{EPD($Graph, node_s, node_d$): }\label{alg:epd}
\begin{algorithmic}
    \If { $Graph$ \textbf{contains} $node_s, node_d$} 
        \State $K_{sd} \gets 0  $
        \State $best \gets [ ]$
        \State $p0 = ShortestPath( Graph, node_S, node_D)$ \footnotemark
        \For{ $path$ \textbf{in} $AllPaths(Graph, node_S, node_D)$} \footnotemark
            
            \If {$path \not= p0$ \textbf{and} $1-\frac{size(p0\oplus path))}{size(path)} > t\_k_{sd}$ }
            \State $best = path$
            \If {$size(best) == 0$} 
                \Comment{Path diversity is 0}
                \State $k_{sd} \gets 0$
            \Else        
            \State $k_{sd} \gets 1-\frac{size(p0\oplus best))}{size(best)} > t\_k_{sd}$ 
            \EndIf
            \EndIf
        \EndFor
        \State \textbf{return} $1-\exp^{-\lambda*K_{sd}}$
    \Else 
    \State \textbf{return} $0$
    \Comment{One or more nodes DNE in network.}
    \EndIf
    
    \small{\Comment{\textsuperscript{1, 2} As both ShortestPath and AllPaths are well-known graph theory algorithms which have numerous possible implementations, we note that any of the commonly-used approaches can be applied to this problem.}}
\end{algorithmic}
\end{algorithm}

For the purpose of our example resilient system graphs, we will utilize the same subsection of physical network, and the same base logical layout, building out connectivity and creating duplication to showcase how TGD can be used to identify maximum resiliency gains for minimal additional infrastructure and complexity. Three values of $\lambda$, \{0.2, 1, 5\}, will be tested to demonstrate the impact they have on resiliency scores as nodes are added and connectivity increased. The base network scenario with limited connectivity, a fully-connected variation, and two variations on the network with a duplicate tracker on the tank level sensor were run through TGD checking, with the results shown in Table 1 below.

\begin{figure}[htbp]
    \centering
    \includegraphics[width=7cm]{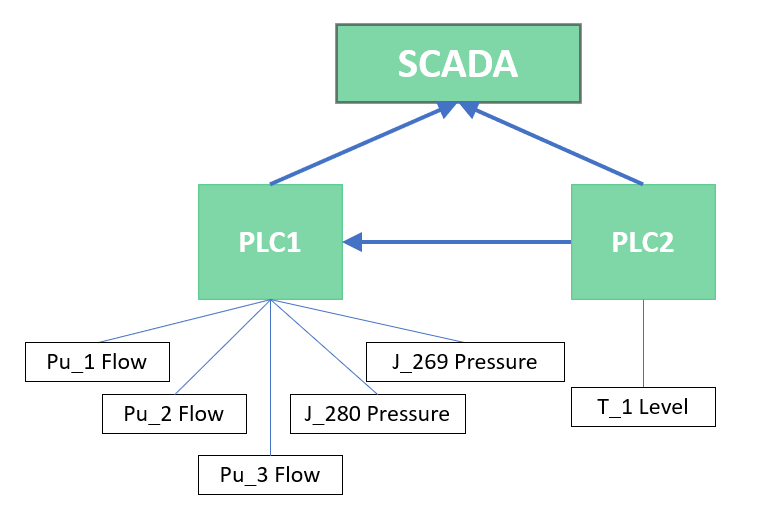}
    \caption{Base Logical Model}
    \label{fig:base_mod}
    \vspace{-10pt}
\end{figure}

\begin{figure}[htbp]
    \centering
    \includegraphics[width=7cm]{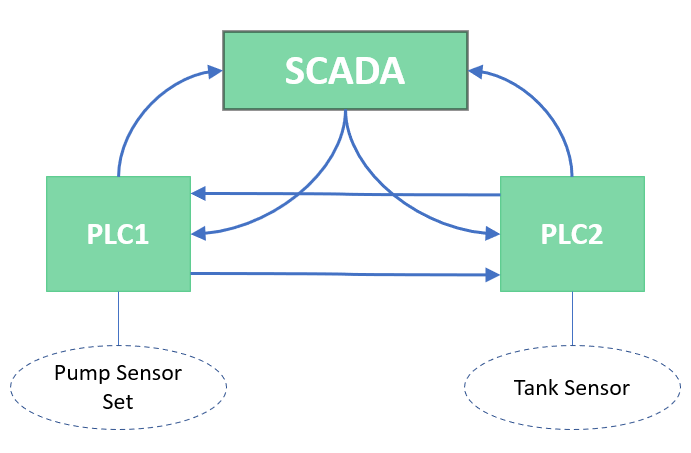}
    \caption{Bidirectional Communication and Checking}
    \label{fig:bid}
    \vspace{-10pt}
\end{figure}

\begin{figure}[htbp]
    \centering
    \includegraphics[width=7cm]{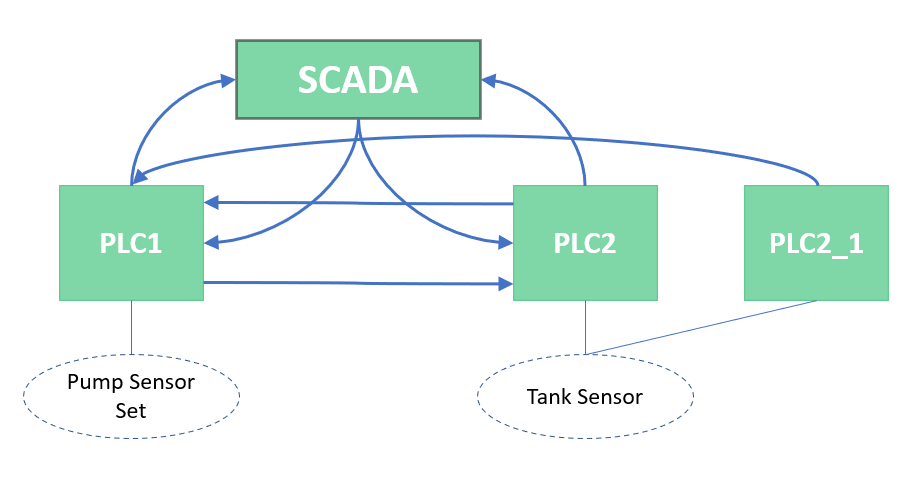}
    \caption{Single Duplication on Tank Level, Partial Connection}
    \label{fig:bid_single_dup}
    \vspace{-10pt}
\end{figure}

\begin{figure}[htbp]
    \centering
    \includegraphics[width=7cm]{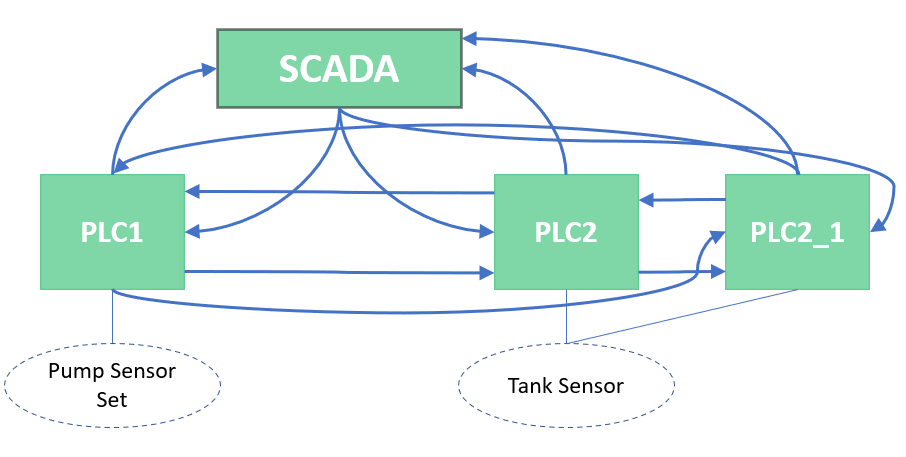}
    \caption{Single Duplication on Tank Level, Full Connection}
    \label{fig:bid_single_full}
    \vspace{-10pt}
\end{figure}

\begin{center}
\label{tbl:lambda}
\begin{tabular}{||c c c c||} 
 \hline
 \multicolumn{4}{|c|}{TGD Values by Topology and $\lambda$} \\
 \hline
 $\lambda$ & 0.2 & 1 & 5 \\ [0.5ex] 
 \hline\hline
 Fig. \ref{fig:base_mod} & 0.0208 & 0.0811 & 0.16072 \\ 
 \hline
 Fig. \ref{fig:bid}  & 0.12482 & 0.48658 & 0.96432 \\
 \hline
 Fig. \ref{fig:bid_single_dup} & 0.08364 & 0.33249 & 0.70274 \\
 \hline
 Fig. \ref{fig:bid_single_full} & 0.09516 & 0.39347 & 0.91791\\ [1ex] 
 \hline
\end{tabular}
\end{center}
You can see that TGD weights heavily on interconnectivity, encouraging two-way communication and checking between cyber nodes. At all $\lambda$ values, it rewards additional connectivity, and penalizes where the number and directness of paths are limited, although additional connectivity is much more heavily rewarded for higher levels of  $\lambda$. It can be seen that simply adding a handful of connections to the base logical model, predicted resiliency (correlated with TGD) can be greatly improved. However, we must also note the severe penalization for the additional duplicate node when said node is not fully connected-- this is not reflective of the likely effect on resiliency, since resiliency should not be negatively impacted by the addition of backup/duplicate sources and checks. This indicates that while TGD can be tested for networks in development and used to estimate relative resilience, additional metrics should be tested and integrated in future releases of \textit{inp2cpa} and any future publications. Nevertheless, resiliency metrics like this are relevant to both standard communication and critical infrastructure control systems networks, and this additionally allows users an initial assessment of the likely resiliency of their network before lengthy simulations and tests are applied.

\section{Conclusion and Future Work}

 The extended functionalities of the tool \textit{inp2cpa} drastically reduces the time it takes to create and run bulk test simulations, as well as enabling technicians and engineers to create custom cyber-physical attacks to simulate on a water network and investigate the impact of the attack. This will be beneficial to engineers and technicians who work with critical infrastructure but do not have the necessary programming experience to efficiently use EPANET and epanetCPA. Future additions will include features and functionality to expand this to include DHALSIM configuration and attack scenario creation, along with a deeper dive into relevant existing resiliency metrics, and potential development of custom resiliency metrics for cyber-physical and smart-device systems based on their unique characteristics. We plan to continue to add to \textit{inp2cpa} to make it as useful as possible to technicians and engineers in the industry. We are currently laying the ground work to add YAML functionality, and are also considering adding python configuration and device file generation, and will consult with members of industry and technical experts to identify what they would see as the most potentially useful additions to any such tool aiming to provide utility and create ease-of-access to simulation and verification toolkits. We believe one of the best ways to move forward with this tool is to eventually migrate it from a guided file creation tool to a fully automated file converter. While some components, such as cyberlinks, cannot be inferred by us and require user input we believe most other functions can be at least in-part, if not fully, automated. Eventually we would also like to see this tool, or core automated components of it, integrated into a simulation toolkit like epanetCPA or DHALSIM. If we were able to integrate this tool with an existing simulation toolkit that would significantly improve the accessibility of that toolkit to technicians in industry and researchers looking to advance knowledge in the field. \textit{inp2cpa}, and tools like it, will make securing critical infrastructure from cyber-physical threats quicker and easier.

\section*{Acknowledgment}
This material is based upon work supported by the National Science Foundation Computer and Information Science and Engineering (CISE) devision, award number 1846493 of the
Secure and Trustworthy Cyberspace (SaTC) program: Formal
TOols foR SafEty aNd. Security of Industrial Control Systems
(FORENSICS).
%To be completed.

\printbibliography[heading=bibliography]

@online{ukraine_2017,
	title = {Ukraine power cut 'was cyber-attack' - {BBC} News},
	url = {https://www.bbc.com/news/technology-38573074},
	urldate = {2021-08-27}
}

@article{mehrpouyan2015resiliency,
  title={Resiliency analysis for complex engineered system design},
  author={Mehrpouyan, Hoda and Haley, Brandon and Dong, Andy and Tumer, Irem Y and Hoyle, Christopher},
  journal={AI EDAM},
  volume={29},
  number={1},
  pages={93--108},
  year={2015},
  publisher={Cambridge University Press}
}

@inproceedings{mehrpouyan2013resilient,
  title={Resilient design of complex engineered systems against cascading failure},
  author={Mehrpouyan, Hoda and Haley, Brandon and Dong, Andy and Tumer, Irem Y and Hoyle, Chris},
  booktitle={ASME International Mechanical Engineering Congress and Exposition},
  volume={56413},
  pages={V012T13A063},
  year={2013},
  organization={American Society of Mechanical Engineers}
}

@inproceedings{mehrpouyan2012model,
  title={A model-based failure identification and propagation framework for conceptual design of complex systems},
  author={Mehrpouyan, Hoda and Jensen, David C and Hoyle, Christopher and Tumer, Irem Y and Kurtoglu, Tolga},
  booktitle={International Design Engineering Technical Conferences and Computers and Information in Engineering Conference},
  volume={45011},
  pages={1087--1096},
  year={2012},
  organization={American Society of Mechanical Engineers}
}

@inproceedings{murillo2020co,
  title={Co-Simulating Physical Processes and Network Data for High-Fidelity Cyber-Security Experiments},
  author={Murillo, Andres and Taormina, Riccardo and Tippenhauer, Nils and Galelli, Stefano},
  booktitle={Sixth Annual Industrial Control System Security (ICSS) Workshop},
  pages={13--20},
  year={2020}
}

@online{futureiot_triton,
	title = {Triton 2.0 and the future of {OT} cyberattacks},
	url = {https://futureiot.tech/whitepaper/triton-2-0-and-the-future-of-ot-cyberattacks/},
	titleaddon = {{FutureIoT}},
	urldate = {2020-08-05},
	langid = {american},
	note = {Library Catalog: futureiot.tech}
}

@inproceedings{cupcarbon,
author = {Bounceur, Ahc\`{e}ne},
title = {CupCarbon: A New Platform for Designing and Simulating Smart-City and IoT Wireless Sensor Networks (SCI-WSN)},
year = {2016},
isbn = {9781450340632},
publisher = {Association for Computing Machinery},
address = {New York, NY, USA},
url = {https://doi.org/10.1145/2896387.2900336},
doi = {10.1145/2896387.2900336},
booktitle = {Proceedings of the International Conference on Internet of Things and Cloud Computing},
articleno = {1},
numpages = {1},
location = {Cambridge, United Kingdom},
series = {ICC '16}
}

@misc{sumo_bisw_international,
	title = {The International Journal on Advances in Systems and Measurements is published by {IARIA}. {ISSN}: 1942-261x},
	shorttitle = {The International Journal on Advances in Systems and Measurements is published by {IARIA}. {ISSN}},
	author = {Bisw, Karabi}
}

@article{abrams_malicious_maroochy,
	title = {Malicious Control System Cyber Security Attack Case Study–Maroochy Water Services, Australia},
	pages = {16},
	author = {Abrams, Marshall D},
	langid = {english}
}

@article{taormina2017characterizing,
  title={Characterizing cyber-physical attacks on water distribution systems},
  author={Taormina, Riccardo and Galelli, Stefano and Tippenhauer, Nils Ole and Salomons, Elad and Ostfeld, Avi},
  journal={Journal of Water Resources Planning and Management},
  volume={143},
  number={5},
  pages={04017009},
  year={2017},
  publisher={American Society of Civil Engineers}
}

@article{TAORMINA201946,
title = {A toolbox for assessing the impacts of cyber-physical attacks on water distribution systems},
journal = {Environmental Modelling \& Software},
volume = {112},
pages = {46-51},
year = {2019},
issn = {1364-8152},
doi = {https://doi.org/10.1016/j.envsoft.2018.11.008},
author = {R. Taormina and S. Galelli and H.C. Douglas and N.O. Tippenhauer and E. Salomons and A. Ostfeld}
}

@online{inp2cpa_source, title = {Nikolopoulos, D. (2020). inp2cpa. In STOP-IT},  
url = {http://tl.stop-it-project.eu/d/Tool/4}}

@article{klise2018overview,
  title={An Overview of the Water Network Tool for Resilience (WNTR).},
  author={Klise, Katherine A and Murray, Regan and Haxton, Terra},
  year={2018},
  publisher={Sandia National Lab.(SNL-NM), Albuquerque, NM (United States)}
}

@article{nikolopoulos_cyber-physical_2020,
	title = {Cyber-Physical Stress-Testing Platform for Water Distribution Networks},
	volume = {146},
	issn = {0733-9372, 1943-7870},
	url = {http://ascelibrary.org/doi/10.1061/%28ASCE%29EE.1943-7870.0001722},
	doi = {10.1061/(ASCE)EE.1943-7870.0001722},
	pages = {04020061},
	number = {7},
	journaltitle = {Journal of Environmental Engineering},
	shortjournal = {J. Environ. Eng.},
	author = {Nikolopoulos, Dionysios and Moraitis, Georgios and Bouziotas, Dimitrios and Lykou, Archontia and Karavokiros, George and Makropoulos, Christos},
	urldate = {2020-07-17},
	date = {2020-07},
	langid = {english}
}

@article{carron2011high,
  title={High-fidelity simulation in the nonmedical domain: practices and potential transferable competencies for the medical field},
  author={Carron, Pierre-Nicolas and Trueb, Lionel and Yersin, Bertrand},
  journal={Advances in medical education and practice},
  volume={2},
  pages={149},
  year={2011},
  publisher={Dove Press}
}

@inproceedings{a0958ff56e4f442ba1a114fd3db062f0,
title = "A Research Agenda in Maritime Crew Resource Management.",
author = "Michael Barnett and David Gatfield and Claire Pekcan",
note = "Publisher: Embry-Riddle Aeronautical University.",
year = "2003",
language = "English",
booktitle = "Proceedings of the International Conference on Team Resource Management in the 21st Century",
publisher = "Embry-Riddle Aeronautical University",
address = "United States",

}

@misc{helmreich2001improving,
  title={Improving teamwork in Organizations},
  author={Helmreich, RL and Wilhelm, JA and Klinect, JR and Merritt, AC},
  year={2001},
  publisher={Hillsdale, NJ: Erlbaum}
}

@article{HETHERINGTON2006401,
title = {Safety in shipping: The human element},
journal = {Journal of Safety Research},
volume = {37},
number = {4},
pages = {401-411},
year = {2006},
issn = {0022-4375},
doi = {https://doi.org/10.1016/j.jsr.2006.04.007},
url = {https://www.sciencedirect.com/science/article/pii/S0022437506000818},
author = {Catherine Hetherington and Rhona Flin and Kathryn Mearns}
}

@article{https://doi.org/10.1111/nicc.12030,
author = {Alinier, Guillaume and Platt, Alan},
title = {International overview of high-level simulation education initiatives in relation to critical care},
journal = {Nursing in Critical Care},
volume = {19},
number = {1},
pages = {42-49},
doi = {https://doi.org/10.1111/nicc.12030},
url = {https://onlinelibrary.wiley.com/doi/abs/10.1111/nicc.12030},
eprint = {https://onlinelibrary.wiley.com/doi/pdf/10.1111/nicc.12030},
year = {2014}
}

@online{dragos_year_2020,
	title = {Year in Review {\textbar} Dragos},
	url = {https://www.dragos.com/year-in-review/},
	urldate = {2021-08-10},
	langid = {american}
}

@article{slowik_evolution,
	title = {Evolution of {ICS} Attacks and the Prospects for Future Disruptive Events},
	pages = {15},
	author = {Slowik, Joseph},
	langid = {english}
}

@online{jun_7_colonial_2021,
	title = {Colonial Pipeline Cyberattack: Timeline and Ransomware Attack Recovery Details},
	url = {https://www.msspalert.com/cybersecurity-breaches-and-attacks/ransomware/colonial-pipeline-investigation/},
	shorttitle = {Colonial Pipeline Cyberattack},
	titleaddon = {{MSSP} Alert},
	author = {Jun 7, Joe Panettieri • and {2021}},
	urldate = {2021-09-26},
	date = {2021-06-08},
	langid = {american}
}

@online{muckrock_aurora,
	title = {Aurora: Homeland Security's secret project to change how we think about cybersecurity},
	url = {https://www.muckrock.com/news/archives/2016/nov/14/aurora-generator-test-homeland-security/},
	shorttitle = {Aurora},
	titleaddon = {{MuckRock}},
	urldate = {2021-09-26},
	langid = {american}
}

@online{cimpanu_two_israel,
	title = {Two more cyber-attacks hit Israel's water system},
	url = {https://www.zdnet.com/article/two-more-cyber-attacks-hit-israels-water-system/},
	titleaddon = {{ZDNet}},
	author = {Cimpanu, Catalin},
	urldate = {2020-07-22},
	langid = {english},
	note = {Library Catalog: www.zdnet.com}
}

@report{hemsley_history_2018,
	title = {History of Industrial Control System Cyber Incidents},
	url = {http://www.osti.gov/servlets/purl/1505628/},
	pages = {INL/CON--18--44411--Rev002, 1505628},
	number = {{INL}/{CON}--18-44411-Rev002, 1505628},
	author = {Hemsley, Kevin E. and E. Fisher, Dr. Ronald},
	urldate = {2020-07-02},
	date = {2018-12-31},
	langid = {english},
	doi = {10.2172/1505628}
}

@inproceedings{murillo_co-simulating_2020,
	location = {Austin {TX} {USA}},
	title = {Co-Simulating Physical Processes and Network Data for High-Fidelity Cyber-Security Experiments},
	isbn = {978-1-4503-9002-6},
	url = {https://dl.acm.org/doi/10.1145/3442144.3442147},
	doi = {10.1145/3442144.3442147},
	eventtitle = {{ICSS} 2020: Sixth Annual Industrial Control System Security Workshop},
	pages = {13--20},
	booktitle = {Sixth Annual Industrial Control System Security ({ICSS}) Workshop},
	publisher = {{ACM}},
	author = {Murillo, Andres and Taormina, Riccardo and Tippenhauer, Nils and Galelli, Stefano},
	urldate = {2021-05-20},
	date = {2020-12-08},
	langid = {english}}

@online{fedtech_cybersecurity_nodate,
	title = {Cybersecurity Lessons Utilities Can Learn from the Oldsmar Water Plant Hack},
	url = {https://biztechmagazine.com/article/2021/04/cybersecurity-lessons-utilities-can-learn-oldsmar-water-plant-hack},
	urldate = {2021-07-13},
	langid = {english}
}

@inproceedings{alenazi_evaluation_2015,
	title = {Evaluation and comparison of several graph robustness metrics to improve network resilience},
	doi = {10.1109/RNDM.2015.7324302},
	pages = {7--13},
	author = {Alenazi, Mohammed and Sterbenz, James},
	date = {2015-10}
}

@article{rohrer_path_2014,
	title = {Path diversification for future internet end-to-end resilience and survivability},
	volume = {56},
	issn = {1018-4864, 1572-9451},
	url = {http://link.springer.com/10.1007/s11235-013-9818-7},
	doi = {10.1007/s11235-013-9818-7},
	pages = {49--67},
	number = {1},
	journaltitle = {Telecommunication Systems},
	shortjournal = {Telecommun Syst},
	author = {Rohrer, Justin P. and Jabbar, Abdul and Sterbenz, James P. G.},
	urldate = {2021-11-29},
	date = {2014-05},
	langid = {english}
}

\end{document}